\documentclass[a4paper,11pt]{article}
\usepackage{amsthm}
\usepackage{amssymb}
\usepackage{amsmath}
\usepackage{amsfonts}
\usepackage{mathrsfs}
\usepackage{bbm}
\usepackage{graphicx}
\usepackage[all,dvips,arc,frame]{xy}
\usepackage{float}

\newcommand{\id}{\ensuremath{\mathbbm{1}}} 

\newcommand{\Ad}{\mathrm{Ad}}

\newcommand{\beq}{\begin{equation}}
\newcommand{\eeq}{\end{equation}}
\newcommand{\beqn}{\begin{eqnarray}}
\newcommand{\eeqn}{\end{eqnarray}}

\theoremstyle{remark}

\title{}

\def\k{\kappa}

\def\de{\delta}

\def\al{\alpha}

\def\Ga{\Gamma}

\def\be{\begin{equation}}
\def\ee{\end{equation}}
\def\bea{\begin{eqnarray}}
\def\eea{\end{eqnarray}}

\def\C{{\mathbb C}}
\def\D{{\cal D}}

\def\G{{\cal G}}
\def\H{{\cal H}}

\def\R{{\mathbb R}}
\def\S{{\cal S}}

\def\si{\sigma}

\def\l{\lambda}

\def\baral{{\bar{\alpha}}}
\def\barb{{\bar{\beta}}}
\def\barga{{\bar{\gamma}}}

\def\e{\varepsilon}
\def\b{\beta}
\def\ga{\gamma}

\begin{document}

\vspace*{0.5cm}
\begin{center}
{\Large \bf Yang-Baxter $\si$ model: Quantum aspects}

\end{center}

\vspace{0.3cm}

\begin{center}

 R. Squellari${}^{a,b}$ \\

\bigskip

{\it ${}^a$Institut de Math\'ematiques de Luminy,
 \\ 163, Avenue de Luminy, \\ 13288 Marseille, France}
 
 \medskip
  
 ${}^b$ \it Laboratoire de Physique Th\'eorique et des
Hautes Energies\\
{\it CNRS, Unit\'e associ\'ee URA 280}\\
{\it 2 Place Jussieu, F-75251 Paris Cedex 05, France}

\bigskip

\end{center}

\vspace{0.2cm}

\vskip 2.5truecm

\begin{abstract}
We study the quantum properties at one-loop  of the Yang-Baxter $\si$-models introduced by  Klim\v c\'\i k\cite{KS3,KS4}. The proof of the one-loop renormalizability is given, the one-loop renormalization flow is investigated  and the quantum equivalence is studied.
\end{abstract}

\newpage
\section{Introduction}
The Yang-Baxter $\si$-models were first introduced by  Klim\v c\'\i k \cite{KS3,KS4} as a special case, at the classical level, of a non-linear $\si$-model with Poisson-Lie symmetry\cite{KS1,KS2}. Recall that the Poisson-Lie symmetry appears to be the natural generalization of the so-called Abelian $T$-duality\cite{Kikka} and non-Abelian $T$-duality \cite{Ossa,Frid,Frad} of non-linear $\si$-models. In particular, two dynamically equivalent $\si$-models can be obtained at the classical level providing that Poisson-Lie symmetry condition holds. That condition takes a very elegant formulation in the case where the target space is a compact semi-simple Lie group which naturally leads to the concept of the Drinfeld double\cite{Drin}. The Drinfeld double is the $2n$-dimensional linear space where both dynamically equivalent theories live. For the Poisson-Lie $\si$-models, a proof of the one-loop renormalizability and quantum equivalence was given in \cite{Sfet1,Sfet2,SQ2,Sfet3}.\\
We are interested by a special class of classical Poisson-Lie $\si$-models, the Yang-Baxter $\si$-models. Those classical models exhibit the special feature to be both  Poisson-Lie symmetric with respect to the right action of the group on itself and left invariant. Thus, using the right Poisson-Lie symmetry or the left group action leads to two different dual theories. Those two dynamically equivalent dual pair of models live in two non-isomorphic Drinfeld doubles, the cotangent bundle of the Lie group for the left action, and the complexified of the Lie group for the right Poisson-Lie symmetry. Classical properties were investigated in the past and it has been showed that Yang-Baxter $\si$-models are integrable \cite{KS3}. More recently, based on the previous work of Refs.\cite{Kawa1,Kawa2,Kawa3}, authors of Ref.\cite{delduc} proved that they belong to a more general class of integrable $\si$-models. In particular, they showed that the $\e$-deformation parameter of the Poisson-Lie symmetry can be re-interpreted as a classical $q$-deformation of the Poisson-Hopf algebra.\\

If classical properties are well investigated, very little is known about the quantum version of the  Yang-Baxter $\si$-models.  In the case where the Lie group is $SU(2)$, the Yang-Baxter $\si$-model coincides with the anisotropic principal model  which is known to be one-loop renormalizable. This low dimensional result can let us hope a generalization for any Yang-Baxter $\si$-models. However, contrary to the anisotropic principal model, the Yang-Baxter $\si$-models contain a non-vanishing torsion which could potentially gives rise to some difficulties.  On the other hand, another generalization of the anisotropic chiral model, the squashed group models are one-loop renormalizable for a special choice of torsion \cite{Orlando}.\\
Furthermore, the one-loop renormalizability of the Poisson-Lie $\si$-model cannot provide any help here since the proof was established for a theory containing $n^2$ parameters when the Yang-Baxter $\si$-models contain only two: the $\e$ deformation and the coupling constant $t$. At the quantum level, the Yang-Baxter $\si$-models are no more a special case of the Poisson-Lie $\si$ models.
 The main result of this article consists in proving the one-loop renormalizability of Yang-Baxter $\si$-models.\\

The plan of the article is as follows. In Section 2 we introduced the Yang-Baxter $\si$-models on a Lie group and all algebraic tools needed. In section 3, the counter-term of the Yang-Baxter  $\si$-models, i.e. the Ricci tensor, is calculated. Section 4 is dedicated to the proof of the one-loop renormalizability, and the computation of the renormalization flow is done in Section 5. In Section 6, we study the quantum equivalence and we express the Yang-Baxter $\si$-action in terms of the usual one of the Poisson-Lie $\si$-models.  Outlooks take place in Section 7.
\\
\section{Yang-Baxter $\si$-models}
\subsection{The complexified double}
We considered the case of the Yang-Baxter models studied in Ref.\cite{KS3}. In that case the Drinfeld double $D$ can be the complexification of a simple compact and simply-connected Lie group $G$, i.e. $D=G^{\mathbb{C}}$, or the cotangent bundle $T^*G$. Let us consider the case of the complexified Drinfeld double, it turns out that $D=G^\mathbb{C}$ admits the so-called Iwasawa decomposition
\beq
G^\mathbb{C}=GAN=ANG.
\eeq
In particular, if $D=SL(n,\mathbb{C})$, then the group $AN$ can be identified with the group of upper triangular matrices of determinant $1$ and with positive numbers on the diagonal and $G=SU(n)$.\\
Furthermore, the Lie algebra $\D$ turns out to be the complex Lie algebra $\G^\mathbb{C}$, which suggests to use the roots space decomposition of $\G^\mathbb{C}$:
\beq
\G^\mathbb{C}=\H^\mathbb{C}\bigoplus (\oplus_{\alpha\in \Delta} \mathbb{C} E_\alpha),
\eeq
where $\Delta$ is the space of all roots. Consider the Killing-Cartan form $\k$ on $\G^\C$, and let us take an orthonormal basis $H_i$ in the $r$-dimensional Cartan sub-algebra $\H^\mathbb{C}$ of $\G^\mathbb{C}$ with respect to the bilinear form $\k$ on $\G^\mathbb{C}$, i.e:
\beq
\k(H_i,H_j)\equiv \de_{ij}
\eeq
This permits to define a canonical bilinear form on $\H^*$, and more specifically endows the roots space $\Delta \subset \H^*$ with an Euclidean metric, i.e.
$$
(\al,\b)=\delta^{ij}\al_i \b_i, \quad \al_i=\alpha(H_i).
$$
Moreover, the inner product on the roots space part of $\G^\C$ is chosen such as:
\beq
\k(E_\al,E_{-\al})\equiv 1,
\eeq
and to fix the normalization, we impose the following non-linear condition $E_\al=E_{-\al}^\dagger$.
 With all those conventions, the generators of $\G^\mathbb{C}$ verify:
\beqn
&[H_i,E_\alpha]=\alpha_i E_\alpha \quad &[E_\alpha,E_{-\alpha}]=\displaystyle \alpha^i H_i  \nonumber\\
&[H_i,H_j]=0 \quad &[E_\alpha,E_\beta]=N_{\alpha,\beta}E_{\alpha+\beta},\; \alpha+\beta\in \Delta.
\label{commutation}
\eeqn
The structure constants $N_{\alpha,\beta}$ vanish if $\alpha+\beta$ is not a root.\\
Since $\G^\C$ is a Lie algebra, the structure constants verify the Jacobi identity which leads on one hand to:
\beq
N_{\al,\b}=N_{\b,-\al-\b}=N_{-\al-\b,\al},
\eeq
and on the other hand to:
\beq
N_{\al,\b+(k-1)\al}N_{\b+k\al,-\al}+N_{-\al,\b+(k+1)\al}N_{\al,\b+k\al}=-(\alpha,\b+k\al).
\label{bianchi2}
\eeq
 In the non-vanishing case, the structure constants $N_{\alpha,\beta}$ can be calculated from the last relation 
\beq
N_{\alpha,\beta}^2=n(m+1)(\alpha, \alpha),
\label{struct}
\eeq
with $(n,m) \in \mathbb{N}$ such that $\beta+n\alpha$ and $\beta-m\alpha$ are the last roots of the chain containing $\beta$ (see Ref.\cite{Gil} for more details).\\
Since $H_i$ is an orthonormal basis in $\H^\mathbb{C}$, we obtain the relations:
\beq
\displaystyle\sum\limits_{\alpha\in\Delta}\alpha_i\alpha_j=\delta_{ij}, \text{ and} \quad \displaystyle\sum\limits_{\alpha\in\Delta}(\alpha,\alpha)=r.
\eeq
A basis of the compact Lie real form $\G$ of $\G^\mathbb{C}$ can be obtained by the following transformations:
\beq
T_i=iH_i,\quad B_\alpha=\frac{i}{\sqrt{2}}(E_\alpha+E_{-\alpha}) \quad C_\baral=\frac{1}{\sqrt{2}}(E_\alpha-E_{-\alpha}),
\label{real}
\eeq
with $\alpha \in \Delta^+$ (positive roots). With our choice of normalization, the vectors of the basis verify $\k(T_i,T_j)\equiv\k_{ij}=-\delta_{ij}$, $\k(B_\alpha,B_\b)\equiv \k_{\al \b}=-\delta_{\al \b}$, $\k(C_\baral,C_\barb)\equiv\k_{\baral \barb}=-\delta_{\al \b}$ and all others are zero. \medskip \\
Let us define now a $\mathbb{R}$-linear operator $R:\G \to\G$ such that:
\beq
RT_i=0, \quad RB_\alpha=C_\baral, \quad RC_\baral=-B_\alpha,
\eeq
 this operator $R$ is the so-called the Yang-Baxter operator \cite{KS4} which satisfies the following modified Yang-Baxter equation:
\beq
[RA,RB]=R\big([RA,B]+[A,RB]\big)+[A,B], \quad (A,B)\in \G.
\eeq
Let us define the skew-symmetric bracket:
\beq
[A,B]_R=[RA,B]+[A,RB], \quad (A,B)\in \G,
\eeq
which fulfills the Jacobi identity, and defines a new Lie algebra $(\G,[.,.]_R)$. It turns out that this new algebra is nothing but the Lie algebra of the $AN$ group of the Iwasawa decomposition of $G^\C$ and will be denoted $\G_\R$ the dual algebra.
\subsection{The Yang-Baxter action}
We shall now consider the action of the Yang-Baxter $\si$-models\cite{KS4} expressed on the Lie group $G$, which takes the expression:
\beq
\S(g)=-\frac{1}{2t}\int \k\big( g^{-1}\partial_+ g,(\id-\e R)^{-1}g^{-1}\partial_- g \big)d\xi^+d\xi^-,\quad g\in G
\label{action}
\eeq
where $\partial_+=\partial_\tau+\partial_\si$ and $\partial_{-}=\partial_\tau-\partial_\si$, $\id$ is the identity map on $\G$, $t$ is the {\it coupling constant} and $\epsilon$ is the {\it deformation} parameter.\\ 
We can immediately check that the Yang-Baxter models \eqref{action} are left action invariant, $G$ acting on himself. Concerning the right Poisson-Lie symmetry, it is well known that such $\si$-models have to fulfill a zero curvature condition to be Poisson-Lie invariant. Indeed, if we take the following $\G^*$-valued Noether current 1-form $J(g)$:
\beq
J(g)=-(\id +\e R)^{-1} g^{-1}\partial_+ gd\xi^++(\id -\e R)^{-1} g^{-1}\partial_- gd\xi^-,
\eeq
we can easily verify that the fields equations of \eqref{action} are equivalent to the following zero curvature condition:
\beq
\partial_+ J_-(g)-\partial_-J_+(g)+\e[J_-(g),J_+(g)]_R=0.
\eeq
We remark that if the deformation $\e$ vanishes then the action of the group $G$ is an isometry, since the Noether current are closed 1-forms on the world-sheets and the action \eqref{action} coincides with that of the principal chiral $\si$-model.\\
The operator $ (\id-\e R)^{-1}$ on $\G$ can be decomposed in a symmetric part interpreted as a metric $g$ on $G$ and a skew-symmetric part interpreted as a torsion potential $h$ on $G$.
 An attentive study of the action \eqref{action} gives the following expressions for $g$ and $h$: 
\beqn
g&=&\k_{i j}(g^{-1}dg)^i (g^{-1}dg)^j  \nonumber \\
&&+\frac{1}{1+\e^2}\bigg(\k_{\alpha \beta}(g^{-1}dg)^\alpha(g^{-1}dg)^\beta+\k_{\baral \barb}(g^{-1}dg)^\baral(g^{-1}dg)^\barb\bigg),\\
h&=&-\frac{\e}{1+\e^2}(g^{-1}dg)^\alpha \wedge (g^{-1}dg)^\baral \k_{\al \al}.
\eeqn
In order to prove the one-loop renormalizability, we need to calculate the Ricci tensor associated to the manifold $(G,g,h)$.


\section{Counter-term of the Yang-Baxter $\si$-models}

In this paper, for the calculus of the counter-term, we choose the standard approach\cite{Fried} based on the Ricci tensor. This choice provides a clear and an elegant expression of the counter-term in terms of the roots of $G^\C$. However the calculus could have been done by using our formula of \cite{SQ2} for the counter-term in an equivalent way.

\subsection{Geometry with torsion on a Lie group $G$}
Let us consider a pseudo-Riemannian manifold $(G,g)$ as the base of its frame bundle, where $G$ is a compact semi-simple Lie group and $g$ a non-degenerated metric. Moreover, we choose the left Maurer-Cartan form $g^{-1}dg,\; g\in G$ as the basis of 1-forms on $G$, and in that basis the metric coefficients $g_{ab}$ and the torsion components $T_{abc}$ are all constant.\\
On that frame bundle we define a metric connection $\Omega$ with its covariant derivative $D$ such that $Dg=0$. Furthermore, if we define by $d^D$ the exterior covariant derivative, the torsion can be written $T=d^D(g^{-1}dg)$. From these definitions we will obtain the expression of the connection $\Omega$.\medskip \\
{\it Metric connection.}\\
By using the relation $Dg=0$ we obtain:
\beq 
\Omega_{\phantom{a}ac}^sg_{sb}+\Omega^s_{\phantom{b}bc}g_{as}=0.
\eeq
With $g_{ab}$ constant and if we denote $\Omega_{abc}=g_{as}\Omega^s_{\phantom{a}bc}$, the previous relation becomes:
\beq
\Omega_{abc}=-\Omega_{bac}
\label{connection1}
\eeq
Thus the two first indices of the connection $\Omega$ are skew-symmetric. \medskip \\
{\it The torsion.}\\
We said that the torsion verifies $T=d^D(g^{-1}dg)$ or in terms of components:
\beq
T^a=\Omega^a_{\phantom{a}b}\wedge(g^{-1}dg)^b+d(g^{-1}dg)^a.
\eeq
Since $g^{-1}dg$ is the left Maurer-Cartan form on $\G$, we get: $$d(g^{-1}dg)=-(g^{-1}dg)\wedge(g^{-1}dg),$$ on the other hand $T^a$ is the 2-form torsion, i.e. we can write it as: $$T^a=\frac{1}{2}T^a_{\phantom{b}bc}(g^{-1}dg)^b\wedge(g^{-1}dg)^c.$$ Consequently, the components of the torsion are related to the skew-symmetric part of the connection as:
\beq
T^a_{\phantom{a}bc}=\Omega^a_{\phantom{b}cb}-\Omega^a_{\phantom{b}bc}-f^{\phantom{bc} a}_{bc},
\label{connection2}
\eeq
with $f^{\phantom{bc} a}_{bc}$ the structure constants of the Lie algebra $\G$.\\
Note that in the case of the non-linear $\si$-models the torsion is defined by $T=dh$ where $h$ is the 2-form potential torsion, we will exploit that a little further to express the connection for the Yang-Baxter $\si$-models.\medskip \\
{\it The connection.}\\
From the relations \eqref{connection1}\eqref{connection2}, we can find the components of the connection:
\beq
2\Omega_{abc}=(-T_{abc}-T_{bca}+T_{cab})+(f_{abc}-f_{cab}-f_{bca}),
\label{co1}
\eeq
with the conventions $\Omega_{abc}=g_{as}\Omega^s_{\phantom{b}bc}$, $T_{abc}=g_{as}T^s_{\phantom{a}bc}$ and $f_{bca}=f^{\phantom{ab}s}_{bc}g_{sa}$.\\
Let us introduce the Levi-Civita connection $L$ which is in fact the second term of the r.h.s in Eq.\eqref{co1}, and rewrite the connection $\Omega$ for a totally skew-symmetric torsion:
\beq
\Omega_{abc}=L_{abc}-\frac{1}{2}T_{abc}.
\eeq
\medskip \\
{\it The curvature and the Ricci.}\\
By definition the 2-form curvature $F$ fulfills $F=d^D \Omega$, i.e.
\beq
F^a_{\phantom{a}b}=d\Omega^a_{\phantom{a}b}+\Omega^a_{\phantom{a}s}\wedge\Omega^s_{\phantom{a}b}.
\eeq
Moreover, since $\Omega^a_{\phantom{a}b}$ is a 1-form of $G$, $\Omega^a_{\phantom{a}b}=\Omega^a_{\phantom{a}bc}(g^{-1}dg)^c$, we obtain the general expression for the curvature:
\beq
F^a_{\phantom{a}bcd}=\Omega^a_{\phantom{a}sc}\Omega^s_{\phantom{a}bd}-\Omega^a_{\phantom{a}sd}\Omega^s_{\phantom{a}bc}-\Omega^a_{\phantom{a}bs}f^s_{cd}.
\eeq
The Ricci tensor is such that $Ric_{ab}=F^s_{\phantom{a}asb}$ and can be written as:
\beq
Ric_{ab}=-\Omega^{s}_{\phantom{ab}ar}(\Omega^{r}_{\phantom{ab}bs}+f^{r}_{\phantom{ab}bs}).
\label{ricci1}
\eeq
We are now able to decompose the symmetric and skew-symmetric parts of the Ricci tensor in terms of the torsion-less Ricci tensor $Ric^L$ and the torsion $T$ as:
\beqn
&Ric_{(ab)}=Ric_{(ab)}^L+\frac{1}{4}T^r_{\phantom{a}as}T^s_{\phantom{a}br} \\
&Ric_{[ab]}=\frac{1}{2}f_{at}^{\phantom{ab}s}T^t_{\phantom{a}bs}-\frac{1}{2}g_{at}f_{sr}^{\phantom{ab}t}g^{ru} T^s_{\phantom{a}bu}   +\frac{1}{2}g_{st}f_{ar}^{\phantom{ab}t}g^{ru} T^s_{\phantom{a}bu}- (a\leftrightarrow b).
\eeqn
\subsection{Application to Yang-Baxter}
\paragraph{Ricci symmetric part}:\\
Recall that in the case of the Yang-Baxter $\si$-models and with our normalization choice, the metric is given by:
\beq
g_{ij}=-\delta_{ij}, \quad g_{\al \b}= -\frac{1}{1+\e^2} \delta_{\al \b}, \quad  g_{\baral \barb}= -\frac{1}{1+\e^2} \delta_{\al \b}.
\eeq
Let us introduce the bi-invariant connection $\Ga$ on the Lie group $G$, it corresponds to the Levi-Civita connection in the case of a vanishing deformation, i.e. $ \Ga=L(\e=0)$. From the equations \eqref{co1} we can obtain the Levi-Civita coefficients:
\beqn
&&L^\al_{\phantom{a}\baral i}=-L^\baral_{\phantom{a}\al i}=(1-\e^2)\Ga^\al_{\phantom{a}\baral i}\\
&&L^\al_{\phantom{a}i \baral}=-L^\baral_{\phantom{a}i \al} =(1+\e^2)\Ga^\al_{\phantom{a}i \baral},
\eeqn
where we keep the convention for the indices $i \in \H$ and $\al\in \Delta^+$. All others Levi-Civita coefficients are equal to those of the bi-invariant connection $\Ga$. \\We can now express the torsion-less Ricci tensor $Ric^L$  as a deformation of the usual Ricci tensor $Ric^\Ga$  of the bi-invariant connection on Lie group, i.e.
\beqn
Ric^L_{\phantom{a}\al \b}&=&Ric^\Ga_{\al \b}-\frac{\e^2}{2}  (\al,\al) \delta_{\al \b} \\
Ric^L_{\phantom{a}\baral \barb}&=&Ric^\Ga_{\baral \barb}-\frac{\e^2}{2}  (\al,\al) \delta_{\al \b} \\
Ric_{\phantom{a}ij}^L&=&(1+\e^2)^2 Ric^\Ga_{\phantom{a}ij}
\eeqn
It is well-known that for the Riemannian bi-invariant structure the Ricci tensor takes the expression:
\beq
Ric^\Ga_{ab}=-\frac{1}{4} \k_{ab}, \quad (a,b) \in G,
\eeq
therefore, the components of $Ric^L$ are the following:
\beqn
Ric^L_{\phantom{a}\al \b}&=&Ric^L_{\phantom{a}\baral \barb}=-\frac{1}{4} \k_{\al\b}-\frac{\e^2}{2}  (\al,\al)\delta_{\al \b} \\
Ric_{\phantom{a}ij}^L&=&-\frac{1}{4} \k_{ij}(1+\e^2)^2
\eeqn
Concerning the contribution of the Torsion to the symmetric part of the Ricci tensor, we have to express the Torsion in terms of the constant structures of $G$. For a non-linear $\si$-model the Torsion $3$-form is calculated from the potential torsion $2$-form such $T=dh$, which implies that:
\beq
T_{abc}=-3f^{\phantom{ab}s}_{[ab}h_{c]s}, \quad (a,b,c,s) \in G.
\eeq
Moreover, since the torsion potential involves only root indices $$h=-\frac{\e}{1+\e^2}(g^{-1}dg)^\alpha \wedge (g^{-1}dg)^\baral \k_{\al \al},$$ the torsion components vanish for the Cartan sub-algebra indices ($T_{ibc}=0$). \\
We can now calculate the torsion contribution, and we obtain for the non-vanishing coefficients:
\beq
\frac{1}{4}T^r_{\phantom{a}\al s}T^s_{\phantom{a}\al r}=\frac{1}{4}T^r_{\phantom{a}\baral s}T^s_{\phantom{a}\baral r}=\frac{\e^2}{2}\big(\frac{1}{2} \k_{\al \al}+ (\al,\al)\big).
\eeq
In the calculus we used the fact that the Killing $\k$ can be expressed in terms of the root $\al$ and the constant structures $N_{\al,\b}$ such as:
\beq
-\frac{1}{2}\k_{\al \al}=\alpha^i \al_i +\frac{1}{2} \sum_{\b \in \Delta^+} (N_{\al,\b})^2+(N_{\al,-\b})^2.
\label{kill}
\eeq
Adding both contributions to the Ricci tensor and using our normalization, we obtain the final expression of the symmetric part:
\beqn
&&Ric_{\al \b}=Ric_{\baral \barb}=-\frac{\k_{\al \b}}{4}(1-\e^2)=\frac{1}{4}(1-\e^2)\delta_{\al \b} \\
&&Ric_{ij}=-\frac{\k_{ij}}{4}(1+\e^2)^2=\frac{\delta_{ij}}{4}(1+\e^2)^2.
\eeqn
We observe that, in the case of the Yang-Baxter model, the torsion induced by the Poisson-Lie symmetry is precisely that which avoids the dependence of the Ricci tensor in the root length $(\al,\al)$.
\paragraph{Ricci skew-symmetric part}:\\
Using the fact the $T_{iab}=0$, the only non-vanishing non-diagonal components of the Ricci tensor can be written:
\beq
Ric_{\al \baral}=2f_{\al \barb \ga}T_{\baral \barb \ga}\k^{\ga \ga} \k^{\barb \barb}-f_{\baral \b \ga}T_{\al \b \ga}\k^{\ga \ga} \k^{\b \b}-f_{\baral \barb \barga}T_{\al \barb \barga}\k^{\barga \barga} \k^{\barb \barb}.
\eeq
The first r.h.s term can be expressed as a function of the structure constants, $N_{\al,\b}$ such as:
\beq
2f_{\al \barb \ga}T_{\baral \barb \ga}\k^{\ga \ga} \k^{\barb \barb}=2\e\sum_{\b \in \Delta^+}(N_{\al,\b})^2-(N_{\al, -\b})^2 .
\eeq
The two other terms are nothing but the contribution of the roots space (see Eq.\eqref{kill}) to the component $\k_{\baral \baral}$ of the Killing form, i.e.:
\beq
-f_{\baral \b \ga}T_{\al \b \ga}\k^{\ga \ga} \k^{\b \b}-f_{\baral \barb \barga}T_{\al \barb \barga}\k^{\barga \barga} \k^{\barb \barb}=-\frac{\e}{2}\bigg(\k_{\baral \baral}+2 (\al,\al)\bigg).
\eeq 
By summing the Bianchi relations \eqref{bianchi2} on positive roots, we obtain that:
\beq
\sum_{\b \in \Delta^+}(N_{\al,\b})^2-(N_{\al, -\b})^2 =-2(\rho,\al)+(\al,\al),
\eeq
with $$\rho=\frac{1}{2} \sum_{\al \in \Delta^+} \al$$ the Weyl vector.\\
Finally, the skew-symmetric part of the Ricci tensor is given by:
\beq
Ric_{\al \baral}=-Ric_{\baral \al}=-\e\bigg(2(\al,\rho)+\frac{1}{2} \k_{\al \al}\bigg).
\eeq

\section{One-loop renormalizability}
At one-loop the counter-terms for a non-linear $\si$-model\cite{Fried} on $G$ are given by:
\beq
\frac{1}{4\pi\epsilon}\int Ric_{ab}(g^{-1}\partial_-g)^a(g^{-1}\partial_+g)^b, \quad \epsilon = 2-d.
\eeq
We require, for the renormalizability, that all divergences have to be absorbed by fields-independent  deformations of the parameters $(t,\e)$ and a possible non-linear fields renormalization of the fields $(g^{-1}\partial_\pm g)^a$. Thus, if we suppose that all parameters  are the independent coupling constants of the theory, the Ricci tensor in our frame has to verify the relations: 
\beqn
Ric_{ab}=-\chi_0(\id-\e R)^{-1}_{ab}+\chi_\e \frac{\partial}{\partial \e}(\id-\e R)^{-1}_{ab}+D_b u_a,
\eeqn
with $u$ a vector that contributes to the fields renormalization, $\chi_0$ and $\chi_\e$ are coordinates-independent. Decomposing into symmetric and skew-symmetric parts, the previous relation for the Yang-Baxter $\si$-models becomes:
\beqn
Ric_{ij}&=&-\chi_0 g_{ij} \label{rel1}\\
Ric_{\al \al}&=& -\chi_0 g_{\al \al}-\chi_\e \frac{2\e}{1+\e^2} g_{\al \al}\label{rel2}\\
Ric_{\al \baral}&=&-\chi_0 h_{\al \baral}+\chi_\e \frac{1-\e^2}{\e(1+\e^2)} h_{\al \baral}+D_{\baral}u_\al.
\label{rel3}
\eeqn
From the equations \eqref{rel1} and \eqref{rel2}, we extract immediately:
\beq
\chi_0=\frac{1}{4}(1+\e^2)^2, \textrm{ and  }\; \chi_\e =-\frac{1}{4}\e(1+\e^2)^2.
\eeq
Since $\chi_0$ and $\chi_\e$ are now fixed, they have to fulfill in the same time the relation \eqref{rel3}, which gives the following constraint:
\beq
\e\bigg(-\frac{1}{2}+2(\rho,\al)\bigg)=-\frac{1}{2} \e + D_{\baral}u_\al.
\label{constraint}
\eeq
Furthermore, the covariant derivative of $u$ can be easily calculated:
\beq
Du=-\frac{1}{2}\sum_{\al \in \Delta^+}(u,\al) (g^{-1}dg)^\al \wedge (g^{-1}dg)^\baral.
\label{cov}
\eeq
Let us define the vector $\e \bar{u}=u$, and insert \eqref{cov} in the constraint \eqref{constraint} we obtain:
\beq
(4\rho-\bar{u},\alpha)=0.
\eeq
Then, if we impose $\bar{u}=4\rho$ the constraint is fulfilled for any root $\alpha$ since $(.,.)$ is the canonical scalar product on $\R^r$ . We can conclude that the Yang-Baxter $\si$-models are one-loop renormalizable.\\
We note that it is quite elegant to find a field renormalization given by the Weyl vector.

\section{Renormalization flow}
\begin{figure}[!here]
\centering
\includegraphics[scale=0.5]{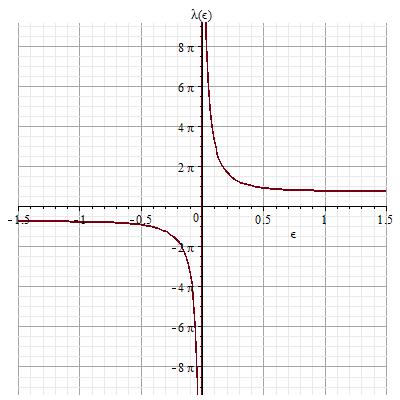}
\caption{\label{fig}Energy scale $\l$ as a function of the deformation parameter $\e$}
\end{figure}

Let us introduce the $\b$-functions of the two parameters $(t,\e)$, they satisfy:
\beq
\b_t=\frac{dt}{d\l}=-t^2 \chi_0, \quad \b_\e=\frac{d\e}{d\l}=t\chi_\e,
\eeq
where $\l=\frac{1}{\pi}\ln \mu$, with $\mu$ the mass energy scale. We obtain the following system of differential equations:
\beqn
\frac{dt}{d\l}+\frac{1}{4}(1+\e^2)^2t^2&=&0\\
\frac{d\e}{d\l}+\frac{1}{4}\e(1+\e^2)^2t&=&0.
\eeqn
The set of differential equations can be exactly solved, and solutions take the following general expressions:
\beq
t(\e)=A \e, \quad \hat{\l}(\e)=B\l(\e)=\frac{3}{2} \arctan \e + \frac{1+ \frac{3}{2}\e^2}{\e(1+\e^2)^2}+C,
\eeq
with $(A,B,C)\in \R$ three integrative constants. We note that divergences occur for $\e$ and $t$ when the energy scale $\hat{\l}$ goes to $\pm\frac{3\pi}{4}+C$. On the other hand, for $\hat{\l}\to \infty$ the parameters $\e$ and $t$ are vanishing, leading to an asymptotic freedom. We can illustrate the situation with the following plot (Fig.\ref{fig}) of $\l$ as a function of $\e$  where we choose $B=1$ and $C=0$.
\newpage

\section{Poisson-Lie models and duality}
Now we will express the Yang-Baxter $\si$-models in terms of the usual Poisson-Lie $\si$-models' expression. Recall that  general {\it Right} symmetric Poisson-Lie $\si$-models can be written:
\beq
 \mathcal{S}(g)=\frac{1}{2\tilde{t}}\int (\partial_- gg^{-1})^a(M+\Pi_R(g))^{-1}_{ab}(\partial_+ gg^{-1})^b.
\label{actionPL}
\eeq
Here $\Pi_R(g)$ is the so-called {\it Right} Poisson-Lie bi-vector and $M$ an $n^2$ real matrix.\\ Using the adjoint action of an element $g\in G$ we can rewrite the action \eqref{action} such as the previous \eqref{actionPL}, with $$\Pi_R(g)=\Ad_g  R \Ad_{g^{-1}}-R\textrm{ and } M=\frac{1}{\e} \id-R.$$
Let us focus on the dual models, as evoked earlier there exists two non-isomorphic Drinfeld doubles for the action \eqref{actionPL}. Consequently, we have two different dual theories for one single initial theory on $G$, and all three are classically equivalent. We will consider each case and argue that they are all quantum-equivalent at one-loop. \medskip \\
We start by considering the Drinfeld double $D=G^\C$, in that case we saw that the dual group is the factor $AN$ in the Iwasawa decomposition. The corresponding algebra is the Lie algebra $\G_\R$ generated by the $\R$-linear operator $(R-i)$ on $\G$, whose its group is a non-compact real form of $G^\C$ (see \cite{KS4,KP} for details). The dual action can be expressed as:
\beq
 \mathcal{S}(\hat{g})=\frac{1}{2\e t}\int (\partial_- \hat{g}\hat{g}^{-1})_a\big[(M^{-1}+\hat{\Pi}_R(\hat{g}))^{-1}\big]^{ab}(\partial_+ \hat{g}\hat{g}^{-1})_b.
\label{dual1}
\eeq
 K.Sfetsos and K.Siampos proved in \cite{Sfet1} that for {\it Right } Poisson-Lie symmetric $\si$-models the quantum equivalence holds providing that the matrix $M$ is invertible. In the Yang-Baxter $\si$-models this condition is always satisfied and the inverse of $M$ is given by: $$M^{-1}=\frac{\e^2}{1+\e^2}\left(\frac{1}{\varepsilon}\,\id+R\right)$$.
\medskip \\
When we consider the dual model associated to the left action of $G$, the Drinfeld double is the cotangent bundle $T^*G=G \ltimes \G^*$. Then the dual group is the dual linear space $\G^*$ of $\G$, which is an Abelian group with the addition of vectors as the group law. The corresponding action is that of the non-Abelian $T$-dual $\si$-models \cite{Ossa,Frid,Frad} and has the well-known expression:
\beq
\mathcal{S}(\hat{g}=e^{s\chi})=\frac{1}{2\e t}\int  d\xi^+d\xi^-\partial_- \chi_a \big((M^{-1})_{ab}+ f_{ab}^c\chi_c\big)^{-1}\partial_+ \chi_b,\; \chi\in \G^*,\; s\in \R.
\label{dual2}
\eeq
It has been showed in \cite{Gal1} that those models are one-loop renormalizable. Since the action \eqref{dual2} is {\it Left } Poisson-Lie symmetric, Sfetsos-Siampos condition \cite{Sfet1} still holds (in their {\it Left} formulation) and implies again the quantum equivalence at one-loop.


\section{Outlooks}
Yang-Baxter $\si$-models are one case of non-trivial Poisson-Lie symmetric $\si$-models which keep the renormalizability and the quantum equivalence at the one-loop level, and are known to be classically integrable. Those models appear to be a semi-classical $q$-deformation of Poisson algebra, and can be a starting point in the quest for a quantum $q$-deformation fully renormalizable thanks to the relative simplicity of these models  containing  only two parameters .\\
Furthermore, for low dimensional compact Lie groups $G$ the geometry associated to the Yang-Baxter $\si$-models can be viewed as a torsionless Einstein-Weyl geometry. We plan in the future to study the Weyl connections with torsion on Einstein manifolds, with the hope to learn more about the geometric aspects of the Poisson-Lie $\si$-models.

\bigskip
I thank G.Valent for discussions and C.Carbone for proofreading.


\end{document}